\documentclass[twocolumn,aps,showpacs,superscriptaddress,amsmath,amsfonts,amssymb,floatfix,prl]{revtex4}
\usepackage{graphicx}
\usepackage{dcolumn}
\usepackage{bm}
\newcommand{\n}{\noindent}

\begin{document}

\title{Entangling power of quantum chaotic evolutions via operator entanglement}

\author{Jayendra N. Bandyopadhyay}
\email{jayendra@prl.ernet.in}
\affiliation{Physical Research Laboratory, Navrangpura, Ahmedabad 380009,
India}
\author{Arul Lakshminarayan}
\email{arul@physics.iitm.ac.in}
\affiliation{Department of Physics, Indian Institute of Technology, Madras, 
Chennai 600036, India}

\begin{abstract}
We study operator entanglement of the quantum chaotic evolutions. This study
shows that properties of the operator entanglement production are 
qualitatively similar to the properties reported in literature about the pure
state entanglement production. This similarity establishes that the operator
entanglement quantifies {\it intrinsic} entangling power of an operator. The
term `intrinsic' suggests that this measure is independent of any specific 
choice of initial states.
\end{abstract}


\maketitle

Entanglement is a unique quantum phenomenon which continues to baffle and 
surprise. Recently, this remarkable property of quantum mechanical 
systems has been identified as a quantum resource whose production is an 
elementary prerequisite for any quantum informational and computational tasks
\cite{nielsen}. This basic task is accomplished by unitary transformations, 
i.e., a given unitary operator acts on a product state and transforms that 
state into an entangled state. Then this given operator is referred to 
as an {\it entangling operator}. However, entangling power of different unitary
operators are naturally not the same. Two different unitary operators 
operating on two identical initial states can produce different entanglement. 

Investigations of different quantum signatures, like spectral statistics of 
the quantum chaotic Hamiltonian, phase space scarring, fidelity decay, etc., of 
classically chaotic systems is the subject of ``quantum chaos" \cite{haake}. 
Recent studies have shown that entanglement in chaotic systems can also be a 
good indicator of the regular to chaotic transition in its classical 
counterpart \cite{furuya,sarkar,arul,our1,tanaka,our2,lahiri,scott,jacquod,
wang,ghose,our3}. Some of these studies have reported that the presence of 
chaos enhances the entanglement production rate \cite{furuya,sarkar}. However, 
in our earlier works, we have observed saturation of entanglement production 
for strongly coupled strongly chaotic systems \cite{arul,our1,our2}. This 
saturation of the entanglement production is a statistical property which we 
have modeled by random matrix theory (RMT) \cite{our1}. Coupling strength 
between two chaotic subsystems is another important parameter for the 
entanglement production. For example, in case of weak coupling, the 
entanglement production is higher for sufficiently long time corresponding to 
nonchaotic cases \cite{our2}.

Recently, rather than focusing on the entanglement evolution of single initial 
product states, the authors studied {\it global} entangling properties of  
coupled chaotic systems \cite{kus} using the coupled kicked tops model
\cite{sarkar,our2}. Following Ref. \cite{zanardi}, they have considered 
different ensembles of initial product states and have studied ensemble 
average of the entanglement production as a measure of entangling power of 
coupled kicked tops time evolution operator $U_T$. Their results have shown 
that the entangling power of $U_T$ is quantitatively, as well as qualitatively 
different for two different ensembles of initial states, one that is averaged 
over all product coherent states, the other including all appropriate product 
states \cite{kus}. Thus this averaging gives a more global measure of 
entanglement.
 
In the present paper, we study {\it operator entanglement} of quantum chaotic
evolution operator. We find that the operator entanglement production behaves 
qualitatively similar to pure state entanglement production \cite{our2}. This 
similarity justifies that the operator entanglement is an intrinsic measure of 
the entangling power of a given operator. The term `intrinsic' is referred to 
the fact that this measure is independent of any specific choice of single 
initial state or some ensemble of initial states. Indeed the operator 
entanglement is only a property of the operator and makes no reference to 
states it acts on. Furthermore, it has been shown in Ref. \cite{wangetal} that 
the operator entanglement is related to the measure proposed in \cite{zanardi}.
However, the operator entanglement is easier to calculate than the measure 
proposed in \cite{zanardi}. 
 
For a formal definition of this measure, let us start with a simple algebraic 
transformation on matrices called {\it matrix reshaping} \cite{karol}. Consider
a rectangular matrix $A$ with elements $A_{ij},\,i\,=\,1,\dots,K$ and 
$j\,=\,1,\dots,L$. This matrix can be reshaped into an one-dimensional vector 
$|A\,\rangle$ by putting its elements row after row into lexicographical order 
of size $KL$, i.e.
\begin{equation}
\begin{split}
a_m = \langle\,m|A\,\rangle &= A_{ij}~~\mbox{where}~~m = (i-1)L + j,\\
i &= 1,\dots,K\,;\,\,j\,=\,1,\dots,L.
\end{split}
\end{equation}   
\n Following is a very simple example of the matrix reshaping \nolinebreak 
\cite{karol} :
\begin{equation}
A = \begin{bmatrix} A_{11} & A_{12}\\A_{21} & A_{22}\end{bmatrix}
\leftrightarrow |A\,\rangle = \{ A_{11}\,\,A_{12}\,\,A_{21}\,\,A_{22}\}^T.
\end{equation}
\n Vector $|A\,\rangle$ can be considered as an element of operator Hilbert 
space or Hilbert-Schmidt (HS) space. The scalar product between any two 
elements $(A, B)$ of a HS space ${\cal H}_{HS}$ is defined as $\langle\,A | 
B\,\rangle \equiv \mbox{Tr} A^{\dagger} B$. Consequently, the HS norm of a 
matrix is equal to just the norm of the associated vector, i.e. $||A||^{2}_{HS}
= \langle\,A | A\,\rangle = \sum_m |a_m|^2$.

Let us now consider an arbitrary unitary operator $U$ operating on a bipartite 
state space ${\cal H} = {\cal H}_1 \otimes {\cal H}_2$, where $\dim 
{\cal H}_1 = N \leq \dim {\cal H}_2 = M$, and $U \in {\cal U}({\cal H})$.
We are interested to measure entangling power of the unitary operator $U$. This
unitary operator can be expanded in terms of complete orthonormal operator 
basis states $\{ |A_m\,\rangle\,\otimes\,|B_\alpha \,\rangle\}$ as
\begin{equation}
|U\,\rangle = \sum_{m=1}^{N^2}\sum_{\alpha=1}^{M^2}  X_{m\alpha}\,| A_m\,
\rangle\,\otimes\,|B_\alpha\,\rangle,
\end{equation}
\n where $|U\,\rangle,\,| A_m\,\rangle$ and $|B_\alpha\,\rangle$ are the 
associated reshaped vectors of the matrices $U,\,A_m$ and $B_\alpha$. The 
operator 
basis states $\bigl\{| A_m\,\rangle\bigr\}$ and $\bigl\{| B_\alpha\,\rangle
\bigr\}$ are orthonormal in the sense that they satisfy $\langle\, A_m | A_n\,
\rangle = \mbox{Tr}(A_m^\dagger\,A_n) = \delta_{mn}$ and $\langle\, 
B_\alpha | B_\beta\,\rangle = \mbox{Tr} (B_\alpha^\dagger\,B_\beta) =
\delta_{\alpha\beta}$. A very simple example of a complete orthonormal operator
basis is $\bigl\{I_2, \sigma_i\bigr\}/\sqrt{2}$, where $I_2$ is the $2 \times 2$
unit matrix and $\sigma_i$'s are the Pauli spin matrices. Now $|U\,\rangle$ 
can be considered as a vector in the composite HS space, ${\cal H}_{HS} 
\otimes {\cal H}_{HS}$. We can apply Schmidt decomposition to $|U\,\rangle$ 
and get
\begin{equation}
|U\,\rangle = \sum_{m=1}^{N^2} \sqrt{\lambda_m}\,|\widetilde{A}_m\,\rangle\,
\otimes\,|\widetilde{B}_m\,\rangle,
\end{equation}
\n where $\{\lambda_m\}$ are the singular values of the rectangular matrix
$X$, and $\bigl\{ |\widetilde{A}_m\,\rangle\bigr\}$ and $\bigl\{ 
|\widetilde{B}_m\,\rangle\bigr\}$ are the new orthonormal basis states.
$\{\lambda_m\}$ can also be identified as the nonzero eigenvalues of operator 
reduced 
density matrices. The operator Schmidt decomposition has also been proposed by 
Nielsen {\it et al} \cite{nielsenetal}. However, application of the matrix 
reshaping operation has put operator Schmidt decomposition and state Schmidt 
decomposition on a same footing. Here we notice $\langle\,U | U\,\rangle = 
\sum_{m=1}^{N^2}\,\lambda_m = \mbox{Tr}(U^\dagger U) = N M$, i.e. the 
vector $| U\,\rangle$ is not normalized. To normalize $|U\,\rangle$, we define 
$\widetilde{\lambda}_m \equiv \lambda_m/N M$. We now define von Neumann entropy
$S_V(U)$ and linear entropy $S_L(U)$ of the operator entanglement respectively 
as
\begin{eqnarray}
S_V(U) &\equiv&  -\sum_{m=1}^{N^2} \widetilde{\lambda}_m\,\ln 
\widetilde{\lambda}_m = -\sum_{m=1}^{N^2} \frac{\lambda_m}{N M}
\ln \frac{\lambda_m}{N M}\\
S_L(U) &\equiv& 1 - \sum_{m=1}^{N^2} \widetilde{\lambda}_m^2 =
1 - \sum_{m=1}^{N^2} \frac{\lambda_m^2}{N^2 M^2}.
\end{eqnarray}
\n These measures have already been utilized to study entanglement 
capability of qudit gates \cite{wangetal}. Both the measures give qualitatively
similar results, but the von Neumann entropy is a more acceptable measure. 
Therefore, in the present paper we prefer $S_V(U)$ to investigate the 
entangling power of quantum chaotic unitary operators.

We use coupled kicked tops as our model of coupled chaotic system \cite{sarkar,
our2}. The time evolution operator, defined in between two consecutive kicks, 
corresponding to the coupled kicked tops is given by
\begin{equation}
U_T = U_{12}^{\epsilon} (U_1 \otimes U_2) =  U_{12}^{\epsilon} \bigl[
(U_1^k U_1^f) \otimes (U_2^k U_2^f) \bigr]
\end{equation}
\n where the different terms are given by,
\begin{equation}
\begin{split}
U_i^f \equiv \exp\left(-i\frac{\pi}{2} J_{y_i}\right),~~ & U_i^k \equiv 
\exp\left(-i\frac{k}{2j_i} J_{z_i}^2\right),\\
U_{12}^{\epsilon} \equiv \exp & \left( -i\frac{\epsilon}{\sqrt{j_1 j_2}} 
J_{z_1} J_{z_2}\right) 
\end{split}
\end{equation}
\n and $i = 1, 2$ represents two different tops. The term $U_i^f$ describes 
free precession of the top around $y$ axis with angular frequency 
$\pi/2$, $U_i^k$ represents a torsion about $z$ axis by an angle
proportional to $J_z$ with the proportionality factor $k/2j$, and 
$U_{12}^{\epsilon}$ is the coupling between the tops using spin-spin 
interaction term with a coupling strength of $\epsilon/\sqrt{j_1 j_2}$.  

Here we study the evolution of the operator entanglement of $U_T$. More 
precisely, we study the operator entanglement of $U_T^n$ as a function of the
time step $n$. In our calculation, we choose complete orthonormal operator 
bases corresponding to each subsystem as,
\begin{equation}
A_\alpha = |m_1\,\rangle\,\langle\,n_1 | ~~ \mbox{and} ~~ B_\beta = |m_2\,
\rangle\,\langle\,n_2 |    
\end{equation}
\n where 
\[\begin{split}
\alpha &\equiv N(m_1 + j_1) + (n_1 + j_1 + 1);\,(m_1, n_1) = -j_1,\dots,j_1\\
\beta &\equiv M(m_2 + j_2) + (n_2 + j_2 + 1);\,(m_2, n_2) = -j_2,\dots,j_2
\end{split}
\]
\n and $N = 2 j_1 + 1,\,\,M = 2 j_2 + 1$. Applying matrix reshaping operation
to $U_T^n$ and to the orthonormal basis states, we can expand $|U_T^n\,\rangle$
as,
\begin{equation}
|U_T^n\,\rangle = \sum_{\alpha=1}^{N^2}\,\sum_{\beta=1}^{M^2}\,u_{\alpha\beta}
(n)\,|A_\alpha\,\rangle \otimes |B_\beta\,\rangle.  
\end{equation}
\n Following the procedure discussed above, we investigate the operator von 
Neumann entropy $S_V(U_T^n)$ as a function of $n$. This study will show how
entangling power of chaotic evolution changes with time steps.

\begin{figure}
\centerline{\includegraphics[height=6.25cm,width=8cm]{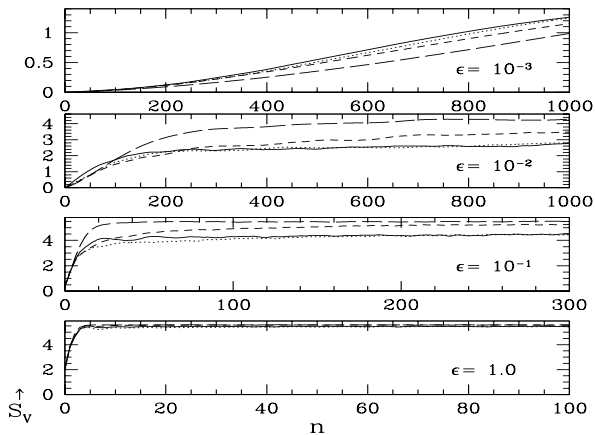}}
\caption{\label{fig1}Time evolution of the operator von Neumann entropy
corresponding to the coupled kicked tops is presented for different coupling
strengths and for different underlying classical dynamics. Solid line
represents $k=1.0$, dotted line corresponds to $k=2.0$, dashed line is for
$k=3.0$ and dash-dot line represents $k=6.0$.}
\end{figure}

In Fig.\ref{fig1}, we have presented our results for the operator 
entanglement production in coupled kicked tops for the spin $j_1 = j_2 = j\,
\mbox{(say)}\,= 10$. As we go from top to bottom windows, coupling strength
is increasing by a factor of ten ($\epsilon = 10^{-4}$ to $\epsilon = 1.0$). 
For each coupling strength, we have studied operator entanglement production
for four different single top kicked strengths $k$, whose corresponding 
classical phase space picture has been presented in Ref.\cite{our2}. 
Here we are presenting a brief qualitative description of those phase space 
pictures to correlate the effect of underlying classical dynamics on the 
operator entanglement production. For $k = 1.0$, the phase space was mostly 
covered by regular orbits, without any visible stochastic region. For $k=2.0$, 
most of the phase space was covered by the regular region, but with a thin 
stochastic layer at the separatrix. For the change in the parameter value from 
$k=2.0$ to $k=3.0$, there was significant change in the phase space. At 
$k=3.0$, the phase space was of truly mixed type. However, the size of the 
chaotic region was much larger than the regular region. Finally, when $k=6.0$, 
the phase space was mostly covered by the chaotic region, with few very tiny 
regular islands.  

Let us first discuss the case of weaker coupling $\epsilon=10^{-3}$, whose
results are presented in the top most window of Fig.\ref{fig1}. Here we observe
larger operator entanglement production for the nonchaotic cases than the 
chaotic one. Therefore, for very weak coupling, the presence of chaos actually 
suppresses the entangling power of $U_T^n$. This property has already been 
observed in \cite{our2,kus}. Present observation shows that the suppression 
of entanglement by chaos is an intrinsic property of coupled chaotic systems 
and is independent of any specific choice of initial state. 

The results corresponding to $\epsilon=10^{-2}$ are presented in the second 
window from the top of Fig.\ref{fig1}. In this case, at least for initial
time steps $(n \lesssim 100)$, we have observed suppression of operator 
entanglement by chaos. However, for larger $n$, the operator entanglement 
corresponding to both the nonchaotic cases $(k=1.0, 2.0)$ eventually saturates 
at a value lower than the values corresponding to the mixed $(k=3.0)$ and the 
chaotic $(k=6.0)$ cases. Here we also notice that the operator entanglement 
production corresponding to the chaotic case $(k=6.0)$ is always larger than 
the mixed case $(k=3.0)$ 

We now come to a reasonably stronger coupling strength $\epsilon = 10^{-1}$, 
for which results are presented in the third window from the top of 
Fig.\ref{fig1}.
Here again we observe saturation of the operator entanglement production for 
all the cases. However, the operator entanglement production rate is now much 
higher than the previous cases. These saturation values are clearly different 
for nonchaotic, mixed and chaotic cases. For the chaotic case $k = 6.0$, our
numerical estimation shows that the saturation value is approximately 
$\ln\bigl(0.6 N^2 \bigr) = \ln (0.6 \times 441) \simeq 5.57$. On the
other hand, the saturation value corresponding to the mixed case $k = 3.0$ is 
slightly lower than this value. But the saturation values corresponding to 
the nonchaotic cases are distinctly lower than the other two cases.

Finally, we discuss the case of very strong coupling $(\epsilon = 1.0)$ for 
which results are presented in the bottom window of Fig.\ref{fig1}.
Here, due to the strong coupling, the over all coupled system is chaotic 
irrespective of underlying classical dynamics of the individual subsystems. 
Consequently the saturation values of the operator entanglement production are 
almost equal to $\ln\bigl(0.6 N^2 \bigr) \sim 5.57$ for all different values 
of $k$. 

\begin{figure}
\centerline{\includegraphics[height=6.25cm,width=8cm]{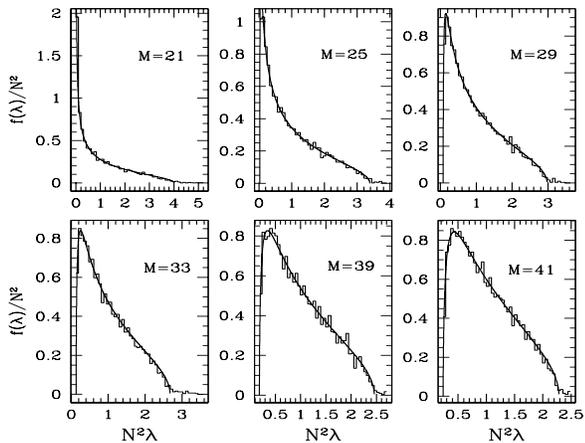}}
\caption{\label{fig2} Distribution of the eigenvalues of the operator RDMs of
coupled kicked tops $(k = 6.0, \epsilon = 1.0)$. Time step $n$ has been taken
in the saturation region. $N = 2 j_1 + 1 = 21$. Solid curves correspond to the
theoretical RMT predicted distribution function.}
\end{figure}

An important outcome of this study is the observation of operator entanglement
saturation at around $\ln(0.6 N^2)$ in case of chaotic subsystems 
which are coupled very strongly. Consequently, from our previous knowledge of 
the saturation of the pure state entanglement \cite{our1}, we expect that the 
distribution of the eigenvalues of the operator reduced density matrices (RDMs)
will follow RMT prediction. In Fig.\ref{fig2}, we have presented the 
distribution of the eigenvalues of the operator RDM corresponding to coupled 
kicked tops for different Hilbert space dimension $M$ of the second top. Here 
we have fixed the dimension of the Hilbert space of the first top at $N = 2j_1 
+ 1 = 21$. The solid curve is representing the RMT predicted Laguerre 
distribution, i.e.,
\begin{equation}
\begin{split}
f(\lambda) &= \frac{N^2 Q}{2\pi} \frac{\sqrt{(\lambda_{\mbox{\small max}} - 
\lambda) (\lambda - \lambda_{\mbox{\small min}})}}{\lambda}\\
\lambda_{\mbox{\small min}}^{\mbox{\small max}} &= \frac{1}{N^2}\left(1 + 
\frac{1}{Q} \pm \frac{2}{\sqrt{Q}}\right)
\end{split}
\end{equation}
where $\lambda \in \bigl[ \lambda_{\mbox{\small min}}, \lambda_{\mbox{\small 
max}} \bigr]$ and $Q = M^2/N^2$ ; and the histograms are the numerical results
obtain from coupled kicked tops model. This figure shows clear agreement
between RMT prediction and numerical results. Using above distribution and  
following the procedure presented in \cite{our1}, we can analytically estimate
the operator entanglement saturation for different dimension ratios $Q$. 
This calculation is straightforward, therefore we do not pursue this further 
here. 

Above studies on the operator entanglement production show some qualitative
properties which are common to the observed properties of the pure state 
entanglement production (see \cite{our2,tanaka,kus}). For instance, (1) the 
operator entanglement, in general, supports `{\it more chaos more entanglement}'
hypothesis; (2) the operator entanglement production has also one statistical 
upper bound which can be explained by RMT, and (3) for the weakly coupled 
cases, chaos suppresses the operator entanglement production. These 
similarities confirm that these are generic intrinsic properties of the 
entanglement production under chaotic evolution. 

Eigenstates of any entangling operator also show some intrinsic entanglement 
properties of that operator. However, in many cases, eigenstates fail to give 
any clue of the entangling power of the operator. Consider an entangling 
operator $U=\exp(-i\alpha J_{z_1} \otimes J_{z_2})$ where $J_{z_i}$'s are usual
angular momentum operators. Its entangling power is determined by the 
parameter $\alpha$. For a definite $\alpha$, this operator can create a 
maximally entangled state in case of a very special initial state. The 
eigenstates of $J_{z_i}$'s, i.e. $\bigl\{ |m_1,m_2\rangle\bigr\}$, are also
the eigenstates of $U$ with eigenvalues $\bigl\{ \exp(-i \alpha m_1 m_2)
\bigr\}$. These eigenstates are all unentangled, but many of them are 
degenerate. Any linear superposition of degenerate eigenstates can form 
entangled eigenstates, which are independent of the parameter $\alpha$. For 
example, $|\phi\rangle = ( |m_1,m_2\rangle + |-m_1,-m_2\rangle)/\sqrt{2}$ are 
the entangled eigenstates of $U$ for $m_1=0$ or $m_2=0$, but $m_1, m_2$ 
are not simultaneously equal to {\it zero}. These eigenstates belong to the 
largest degenerate subspace of dimension $4j+1$ with eigenvalues of 
{\it unity}. However, the maximum possible von Neumann entropy of these 
entangled eigenstates can be $\ln 2$. This is a maximally entangled state only 
for the spin $j=1/2$. Besides, there are some degenerate subspaces with 
dimensions less than $4j+1$, in which we can construct eigenstates with von 
Neumann entropy larger than $\ln 2$ but much less than the maximum possible 
value $\ln(2j+1)$. Above all entanglement of these eigenstates are independent 
of $\alpha$. On the other hand, operator entanglement of $U$ is a function of 
$\alpha$ and it shows qualitatively similar behavior to the pure state 
entanglement production by $U$. In case of the operator $U_T^n$, we have seen 
that its entangling power increases with $n$. But the eigenstates of $U_T^n$ 
are independent of $n$ and consequently it is impossible to distinguish the 
entangling power of $U_T^n$ for different $n$. Let us consider an interesting 
unitary operator $U_p = \exp(-i p J_{z_1}) \otimes \exp(-i p J_{z_2})=
\exp[-i p (J_{z_1} + J_{z_2})]$ whose eigenstates are again $\bigl\{ 
|m_1,m_2\rangle\bigr\}$. This operator is clearly not an entangling operator. 
However, due to degeneracy, there are some eigenstates like $|\phi_p\rangle = 
N^{-\frac{1}{2}} \sum_m |m,-m\rangle$ are maximally entangled. Moreover, from 
different degenerate subspaces, we can construct eigenstates with von Neumann 
entropies $\ln X$ where $X$ can be any real number between $2j+1$ to $1$. 
Therefore, in this case, entanglement of the eigenstates gives an incorrect 
estimation of the entangling power of $U_p$. However, operator entanglement 
of $U_p$ is equal to {\it zero}.

In summary, we have studied operator entanglement production of the chaotic
evolutions. This study shows that the behaviors of the operator entanglement
production are qualitatively similar to the reported behaviors of the pure
state entanglement production by the chaotic evolution operators. Consequently,
this study establishes that the operator entanglement is an intrinsic measure
of the entangling power of an operator. Moreover, we have pointed out that, 
when the eigenstates of a given operator fail to give any clue of its 
entangling power, the operator entanglement can give some estimation. As fur 
as RMT is concerned, we have demonstrated another realization of the Laguerre 
distribution.

\end{document}